\documentclass[runningheads,a4paper]{llncs}

\usepackage{amssymb}
\usepackage{graphicx}
\usepackage{url}
\usepackage{apacite}
\newcommand{\keywords}[1]{\par\addvspace\baselineskip
\noindent\keywordname\enspace\ignorespaces#1}

\begin{document}

\mainmatter  % start of an individual contribution

\title{Translating Paintings Into Music\\Using Neural Networks}

% \titlerunning{Running Title}

\author{Prateek Verma\inst{1,2}\and Constantin Basica\inst{1} \and Pamela Davis Kivelson\inst{3} \thanks{Prateek Verma would like to thank Julius Smith for funding him during the 2020 Spring Quarter at Stanford University. Verma is also an affiliate of the Department of Computer Science.}}
%
% if the names of the authors are too long for the running head, please use the format: AuthorA et al.
%\authorrunning{AuthorA and AuthorB (or AuthorA et al. if too long)}

\institute{Center for Computer Research in Music and Acoustics (CCRMA),\\Stanford University \and Stanford Artificial Intelligence Laboratory \and Design Program, Department of Mechanical Engineering, Stanford University\\\ \email{prateekv@stanford.edu, cobasica@ccrma.stanford.edu, pdk@stanford.edu}}


\maketitle

\begin{abstract}
We propose a system that learns from artistic pairings of music and corresponding album cover art. The goal is to 'translate' paintings into music and, in further stages of development, the converse. We aim to deploy this system as an artistic tool for real time 'translations' between musicians and painters. The system's outputs serve as elements to be employed in a joint live performance of music and painting, or as generative material to be used by the artists as inspiration for their improvisation.
\end{abstract}

\section{Introduction and Related Work}
In our collaboration, we have been exploring how to enable intermedia performance that connects us across distance. Our conversation led us to an experiment asking whether we can render a painting into music and vice versa. Throughout the course of modern history, artists have sought to find connections between musical and visual practices, as well as to construct interrelationships between these art mediums. There exist previous examples of audiovisual encounters that bridge disparate art forms to create hybrid modes of multi-sensory experience. Recently, significant progress has been made in deploying  artificial intelligence to augment the reach of the artists. \cite{gatys2015neural} proposed altering styles of a painting while preserving the content, yielding aesthetically pleasing transformations. Such systems can be assumed to be intelligent automatic accompaniment systems similar to \cite{verma2012real}. Our experiment is different in that the input is both a visual feedback of the content of the image being drawn, and the manner in which the painting is being made via inputs from a contact microphone. In addition, our project looks at how the sound of brush strokes during live painting in tandem with the image itself can be transposed to aesthetically similar music from a given library. 

\section{Discussion and Setup}
As the two artists in our project, respectively a painter (Pamela Davis Kivelson) and a composer (Constantin Basica), we share a formal creative practice built on processes of iteration, layering, and texturing. Another compositional parallel consists of a focus on performance and on vestigial resonances from previously created works. Though our interdependent creativity could seem random, an instinctual pattern emerges during our collaboration. This is developed over the duration of  a joint performance. Spontaneity is also critical during our shared creative experience. Starting from these intersecting congruences in style, we also consider how our mediums support the realization of analogous gestures. Periodicity, rhythm, and intensity are some of the parameters that we can employ to create gestures in music and painting. At the same time, we acknowledge the modal distinction of the two mediums and the limit between what would be considered ‘direct translation’ (e.g., a fast upward brush stroke could be heard as a fast musical scale played upward), which ultimately is less aesthetically interesting, and what would be described as ‘poetic translation’ (e.g., a pattern of dark colors is accompanied by a sequence of dissonant low tones). We mark this distinction through a proposal to introduce a third element in our artistic collaboration. The AI system does not share our artistic patterns and biases, but rather has developed its own. Our hope is that it can introduce an element of surprise in our performance and teach us more about our own creative processes. We envision multiple scenarios for hybrid musical-visual performances employing our AI system. In the following we will describe two scenarios that we plan to undertake in the next phase of our project.

1) The painter and the composer are separated either involuntarily (e.g., as a result of current lockdowns due to the coronavirus pandemic) or voluntarily as an artistic premise. They cannot see or hear each other. The only element that connects them is the AI system. Libraries of extant paintings and compositions of the two artists are implemented in the system prior to the performance. The system listens to the music that is performed live and watches the painting process unfold simultaneously. By measuring the distance between images and sounds in these two processes, the system decides in real time which musical excerpts and which paintings from the pre-established library would fit best with the current input from the two artists. Conversely, the system listens to the music and translates it into one of the paintings from the library, then displays it to the composer. Both artists receive snippets of each other’s preexisting works in real time according to their performance. The audience cannot see this content since its primary function is to offer generative material for the artists. Therefore the AI system achieves a style translation between the painter and composer, offering them material for inspiration to develop further in the performance. An important aspect is that the two artists themselves extend the system’s audiovisual translation by deciding how to continue their musical improvisation and painting based on cross-modal stimuli. In an alternative scenario, the two artists perform together in the same room with the audience. The resulting music based on the live painting is played through speakers and mixed by the composer with their improvisation, while the resulting images based on the live music are mixed into a projection layered with the painter’s work. 
\par
2)	The composer and the painter perform together, remotely or in the same place. They start from a blank slate without discussing any plan. As they begin improvising, the AI system compares their music and painting in real time and decides the level of congruity between them. The artists monitor this level continuously and decide whether they want to remain stylistically aligned or to deliberately stray away from each other. This level of consonance between music and painting may be hidden from the audience or employed as an additional staging element (e.g., it could control the lighting intensity or other multimedia elements).
To gather the data for testing the system, we staged a remote performance with the painter and the composer in which we recorded data from the painter. The composer performed a musical improvisation with digital tools (multiple virtual instruments and effects arranged in the digital audio workstation Ableton Live) and hardware tools (a Roland 88-key MIDI keyboard and a Novation LaunchControl XL MIDI controller). The artist listened to the live music and painted in oil, on several canvases, based on the sonic impressions produced by the music. We want to emphasize that, while we are proposing a translation of painting into music with our AI system, the initial data was gathered through a similar but reversed translation process: from the composer’s live music to the live painting. In this use case the translation is realized by the painter’s artistic sensibility. The artist used a range of traditional and atypical tools to paint with, ranging from big to small brushes, a palette knife and scrapers. There are big differences in brush stiffness and sound making qualities: scraping, erasing, painting over, or superimposing influences the sound of the brush strokes. Comparably, the composer layered his music improvisation by means of loops, delays, feedback, and granular synthesis. Their artistic synthesis was fed to the system, which measured the distance between their styles against two libraries of paintings and musical works supplied by the painter and the composer. The library of paintings consists of three previous works by the artist (Fig. 1), while the musical library is divided in two parts: a list of fifty works by various composers, chosen by Constantin Basica, and a list of forty compositions by himself.

\begin{figure}
	\centering
	\includegraphics[height=3.2cm]{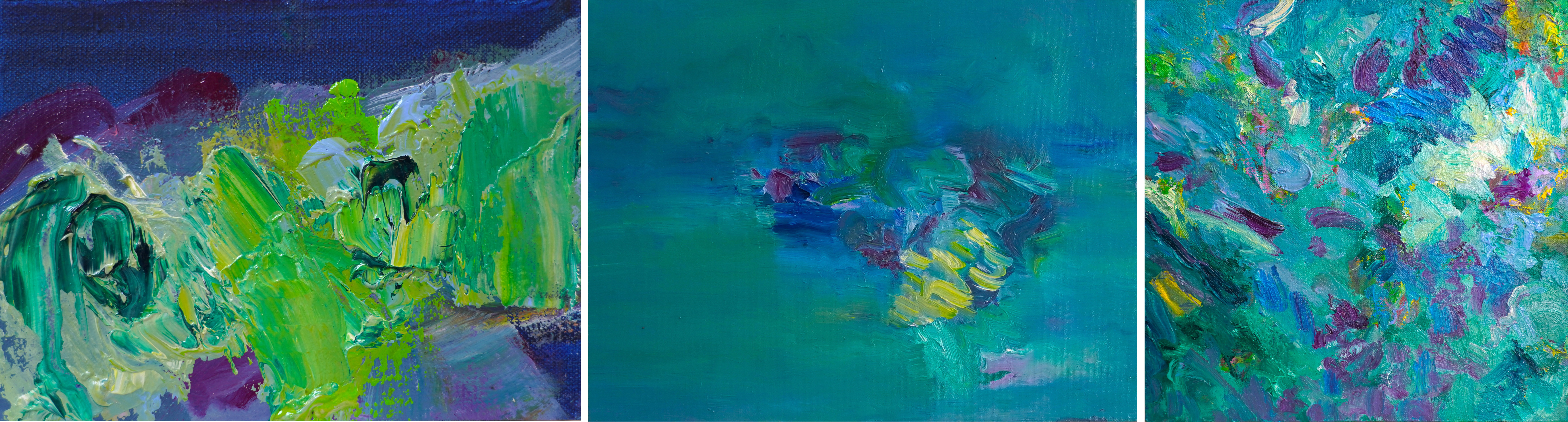}
	\caption{Three paintings realized by Pamela Davis Kivelson during a preparation performance session.}
	\label{fig:paintings}
\end{figure}

For the second filtering level, we employed audio recordings of the brush strokes on canvas measured against the two musical libraries. We used a contact microphone (AKG C411 L) to record these sounds. By utilizing this type of microphone, we could avoid all other surrounding sounds in the room and focus sonically on the brushing of the painter. The microphone was attached to the back of the canvas so as not to interfere with the painting. The contact microphone picked up vibrations produced by the canvas through the excitation of the brush, and transformed them into an audio signal. This was carried to the computer through an audio interface (TASCAM 4x4) and recorded as an audio file. Short snippets of the audio recording were extracted and assigned to the corresponding paintings.

\begin{figure}
	\centering
	\includegraphics[height=3.2cm]{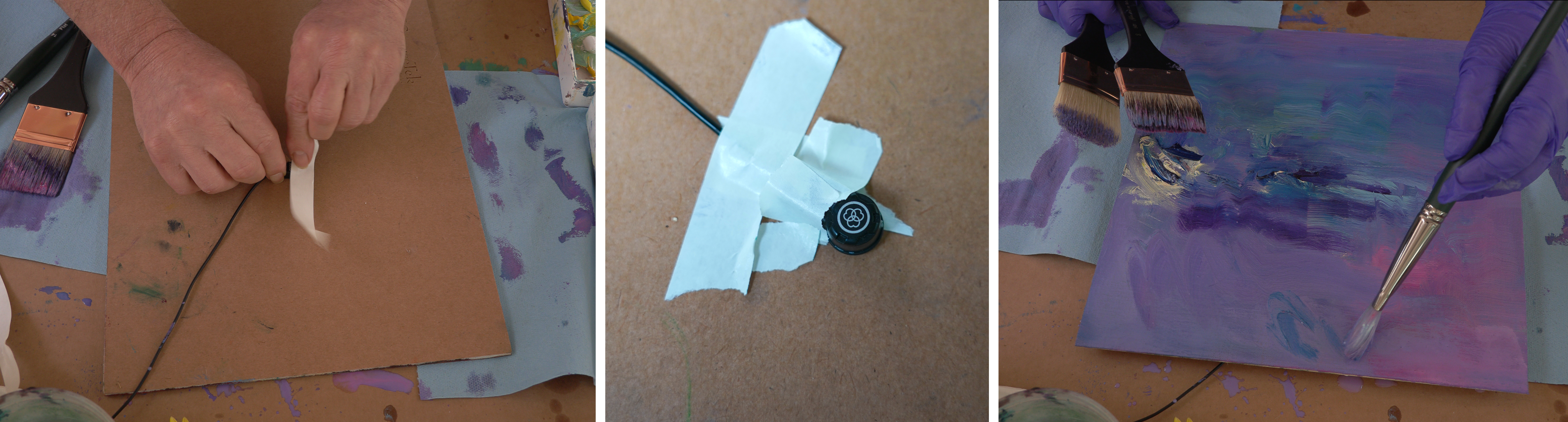}
	\caption{Setup of the contact microphone attached to the back of a canvas by the painter.}
	\label{fig:contact-mic}
\end{figure}

\section{Methodology}
\subsection{Distance Measure between Images and Sounds}
As explained in the previous section, the goal of this work is to translate paintings into music and to develop a tool for performance. Ultimately, we envision giving musicians and artists novel ways to make their performance more interactive through mediums they are familiar with. Specifically, this tool will offer painters the ability to manipulate sound using their paint brushes. In this section, we will explain how we achieved a deep neural network training to accomplish our goal. 

\begin{figure}
	\centering
	\includegraphics[height=7cm]{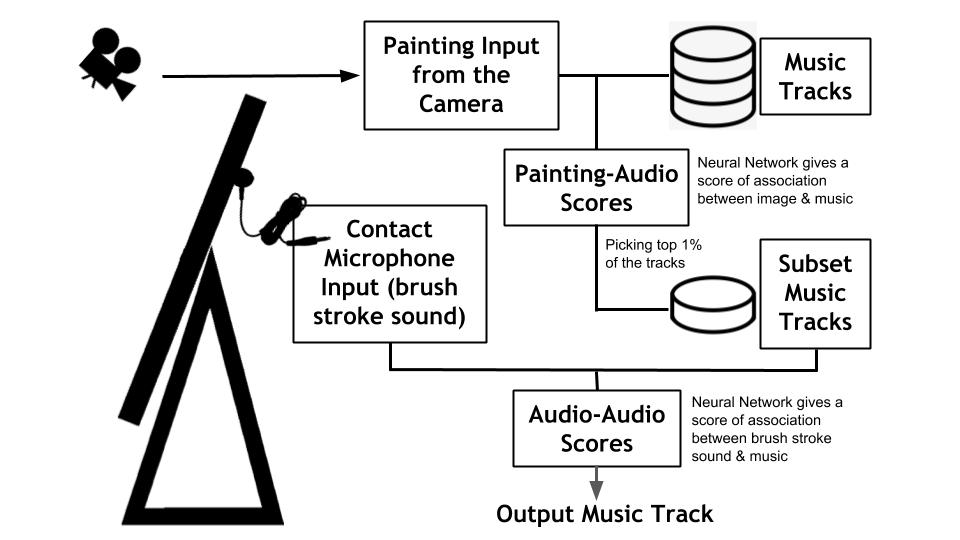}
	\caption{Overview of the system's architecture.}
	\label{fig:diagram}
\end{figure}

The  system involves converting the sounds of a paintbrush striking a canvas and the painting itself in order to select and play out a musical track. This involves drawing a correspondence between visual and sound worlds. In order to associate visual and sound systems, we chose to train a deep neural network to learn to map album cover art with their corresponding sound clip. Our reasoning was that album art contains a diverse representation of elements of the visual world, and a stylistically varied mixture of abstraction and realism. Care is given to each music album’s cover to make the album art stylistically representative of the emotions, contents, and the aesthetics of the music which other sources often do not capture (e.g., a music video). We scraped a publicly available subset of a million song dataset \cite{bertin2011million}, namely MSD-I \cite{oramas2018multimodal}. It contains diverse sets of music pieces along with their respective album cover art. As described in the previous section, we intend to give both artists, namely the musician and the painter, new ways to interact with each other's modality and performance. 
\par
In order to create the dataset, we scraped about 30,000 images along with their corresponding audio tracks from the meta-data available. The architecture follows \cite{arandjelovic2017look} with some modifications. While collecting data, for every positive pair (and assigning a score of 0 equivalent of similar distance) we chose randomly from the track associated with that of the album art. For creation of a negative pair (and assigning a score of 1), we sampled from any of the other tracks except for the track belonging to the current album. Heavy data-augmentation was done for the images (random flips, crops etc.) to create the dataset. For the audio, since we were sampling from anywhere in the track, we introduced random amplitude variations. All of the images were either down-sampled or up-sampled to [224,224,3]. For the audio, we took 4s of audio, downsampling it to 16kHz, and converting it to a mel-represenation with 100 filters and hopping every 12.5ms to give a representation of 100 x 320 for every chunk of audio. ResNet-50 \cite{he2016deep} model was used to project both of these data streams onto 512-dimensional points with two dense layers projecting the concatenation of these modalities to a 2D space. Cross entropy loss was minimized between the softmax activation of the dense layer with that of the ground truth mappings.
After validation on the held-out test set, we obtained an accuracy of about 85\% in categorizing if the system could correctly identify whether the image-audio pair had correspondence or not. The goal of this work was not to compare different architectures to make this even better, but to propose a way to use such architectures for performances, and as tools to artists in mediums they are unfamiliar with. We can now use our model assign scores to any pair of a visual-sound data, with 0 being a strong closeness or association and 1 being dissimilar. Even though the actual test data will not be album cover art or the music it was trained on, the model learns to discern elements from both the music and the art domains, and transfer this skill to the art-music of interest.

\subsection{Distance Measure between Two Sounds}
A contact microphone attached to the back of the painter’s canvas or panel was used to record the physicality in the artist's painting. One can make an analogy between a long passage with varied staccato brush strokes and a violin bowing rapid strokes during fast movements. In order to translate such elements to corresponding musical pieces, we need to learn to map a score to compare any two audio snippets. In cases where there exists a parallel corpus, a neural net can easily learn such mappings as proposed in \cite{haque2018conditional}. We chose to learn how to map such snippets with respect to proximity in a latent space. The audio snippets should understand corresponding correlations in pitch, timbre and rhythm \cite{verma2019neuralogram}. 
We trained DenseNet \cite{huang2017densely} on a balanced AudioSet \cite{audioset} with cross entropy loss on the one-hot categories with the same audio representation i.e. mel-represenation of the audio signals of 4s in duration. This is a state of the art architecture well explored, the details of which lie beyond the scope of this work. After training, we extracted latent representation of the audio signals from the last convolutional layer, as a signature or encapsulation of the contents of the audio  \cite{haque2019audio,verma2019neuralogram}. In order to map the sounds from the contact microphone to that of an actual musical library, we stored all of the embeddings of all the tracks at a 4s resolution. We defined the Euclidean distance between the embeddings of the two audio signals as their distance or a measure of how close two audio signals are. Even if the input is different w.r.t timbral space of the output, still these embeddings will learn one or more aspects of sound like pitch, timbre or rhythm as shown in \cite{verma2019neuralogram}. All of the embeddings were computed and stored so it is easier to retrieve the closest audio to that of the input from the contact microphone almost in real time.

\subsection{Two Step Filtering for Audiovisual Selection}
An embedding based distance metric can associate scores between the visual-audio domains (as learned from album cover art) and audio-audio domains(as learned from an audio understanding model). We now can translate paintings into music, both from the visual and the sound perspectives. The choice of how to combine these two modalities is a design choice, and there are several possible options. One potential filter, for the artist to use while painting, is to choose a fraction of the sounds that are likely to be associated with the painting, i.e. the lowest 1\% of the scores of the sounds that match with that of the painting from the distance metric learned from the album-art trained neural net. This network assigns a score for every pair of painting and sound between 0-1 as to how close-far they are associated. After getting a fraction of our music, which can get associated with the painting, we now search within the fraction of music library to match the sounds of the brush strokes with the musical pieces by finding the closest mapping between the two lists of audio embeddings. 

\subsection{Observations from the Output Analysis}
We see that the triggering from the input of brush strokes—in order to retrieve musical sound—matches elements such as similar onset patterns with same/distinct timbre. The hum and the silent background are also matched with non-onset based elements such as long held notes. Since we filter with the visual aspect before the audio retrieval, our embedding based audio score compares with a diverse set of audio inputs rather than just giving the closest match from the entire dataset of similar audio snippets.

\section{Conclusion and Future Work}
In this project we have proposed a way to augment the work of a painter and a composer by translating paintings into corresponding musical excerpts, which then serve as generative material for the performers. The system was trained on a multi-genre set of music albums from 1927 to 2011. In contrast, when the system was applied in our testing, it looked at and listened to singular, experimental works by two contemporary artists. Regardless of the disparity between the training and the testing material, the system was able to find stylistically interesting correspondences between the artists' practices.
This work has potential in further increasing novel digital representations of performance using AI techniques. As more of our creative activities move online, projects like this could enable geographically displaced performance based on style transfer from one medium to another. We can envision a time when such emerging hybrid style assemblages will allow a more intimate collaboration between artists in different mediums. Our first iteration already allows for new ways to explore interdisciplinary artistic collaboration. Ideally, this system will make visible the unseen dynamics that we feel but do not see.
\par
For a demo based on the works of Pamela Davis Kivelson and Constantin Basica, please visit: 
\url{http://ai.stanford.edu/~prateekv/Demo/}

\bibliographystyle{apacite}
% Place your own .bib file here:
\bibliography{Translating_Painting_Into_Music_Using_Neural_Networks.bib}

\end{document}